# Ultrashort Tilted-Pulse-Front Pulses and Nonparaxial Tilted-Phase-Front Beams

**Liang Jie Wong[1,*] and Ido Kaminer[2,3]**

[1]Singapore Institute of Manufacturing Technology, 2 Fusionopolis Way, Innovis, Singapore 138634

[2]Department of Physics, Massachusetts Institute of Technology, 77 Massachusetts Avenue, Cambridge 02139, Massachusetts, USA

[3]Department of Electrical Engineering, Technion–Israel Institute of Technology, 32000 Haifa, Israel.

*wonglj@simtech.a-star.edu.sg

**Abstract: Electromagnetic pulses with tilted pulse fronts are instrumental in enhancing the efficiency of many light-matter interaction processes, with prominent examples including terahertz generation by optical rectification, dielectric laser acceleration, ultrafast electron imaging and X-ray generation from free electron lasers. Here, we find closed-form expressions for tilted-pulse-front pulses that capture their exact propagation dynamics even in deeply nonparaxial and sub-single-cycle regimes. By studying the zero-bandwidth counterparts of these pulses, we further obtain classes of nondiffracting wavepackets whose phase fronts are tilted with respect to the direction of travel of the intensity peak. The intensity profile of these nonparaxial nondiffracting wavepackets move at a constant velocity that can be much greater than or much less than the speed of light, and can even travel backwards relative to the direction of phase front propagation.**

Tilting the pulse front of an electromagnetic wavepacket with respect to its propagation direction is a technique that has led to significant efficiency enhancements across a wide variety of fields in optics and photonics. In optical rectification, a tilted pulse front facilitates a match between the pump pulse group velocity and terahertz phase velocity, enabling the generation of intense terahertz pulses [1-3]. In optical parametric amplification, a tilted pulse front is used to match signal and idler group velocities during the generation of high-power few-cycle optical pulses [4,5]. Tilting the pulse front has also proven an effective solution to synchronization problems by enabling the prolonged interaction of electron bunch and laser pulse in dielectric laser acceleration [6,7], ultrafast electron diffraction [8] and X-ray generation schemes [9,10]. Although methods of generating tilted-pulse-front pulses are well-known [11,12] and approximate mathematical models for their propagation dynamics have been developed [13,14], a fully-rigorous description of tilted-pulse-front pulses does not currently exist. Such a fully-rigorous description is vital with the continued push towards extreme intensities, extremely short pulse durations and tight focusing, since paraxial approximations breakdown and nothing is known about the properties of tilted-pulse-front pulses in the nonparaxial or few-cycle regimes.

Here, we present exact, fully closed-form expressions for the electromagnetic fields of tilted-pulse-front pulses, which encompass conditions of extreme nonparaxiality and extremely short pulse duration. Associated with these pulses are their zero-bandwidth counterparts, which are fascinating in their own right as new classes of nonparaxial nondiffracting electromagnetic modes. The intensity envelopes of these wavepackets move at a constant velocity that can be much greater than or much less than the speed of light, and generally travel at an angle (possibly backwards) with respect to the direction of phase front propagation.

These new modes add to the diversity of nondiffracting wavepacket solutions, which include the Bessel beam [15], propagation-invariant superpositions of Bessel beams [16,17], X-wave pulses [18,19], subluminal and superluminal wavepackets [20-24], as well as accelerating [25-29] and non-accelerating [30] Airy-shaped pulses. In addition to their fascinating physics, nondiffracting wavepackets have applications that range from materials processing to medical care [31-34]. The new electromagnetic modes we present are closed-form solutions to Maxwell's equations, which makes them attractive candidates for modeling light-matter interactions ranging from particle acceleration and X-ray generation to optical parametric amplification and materials processing.

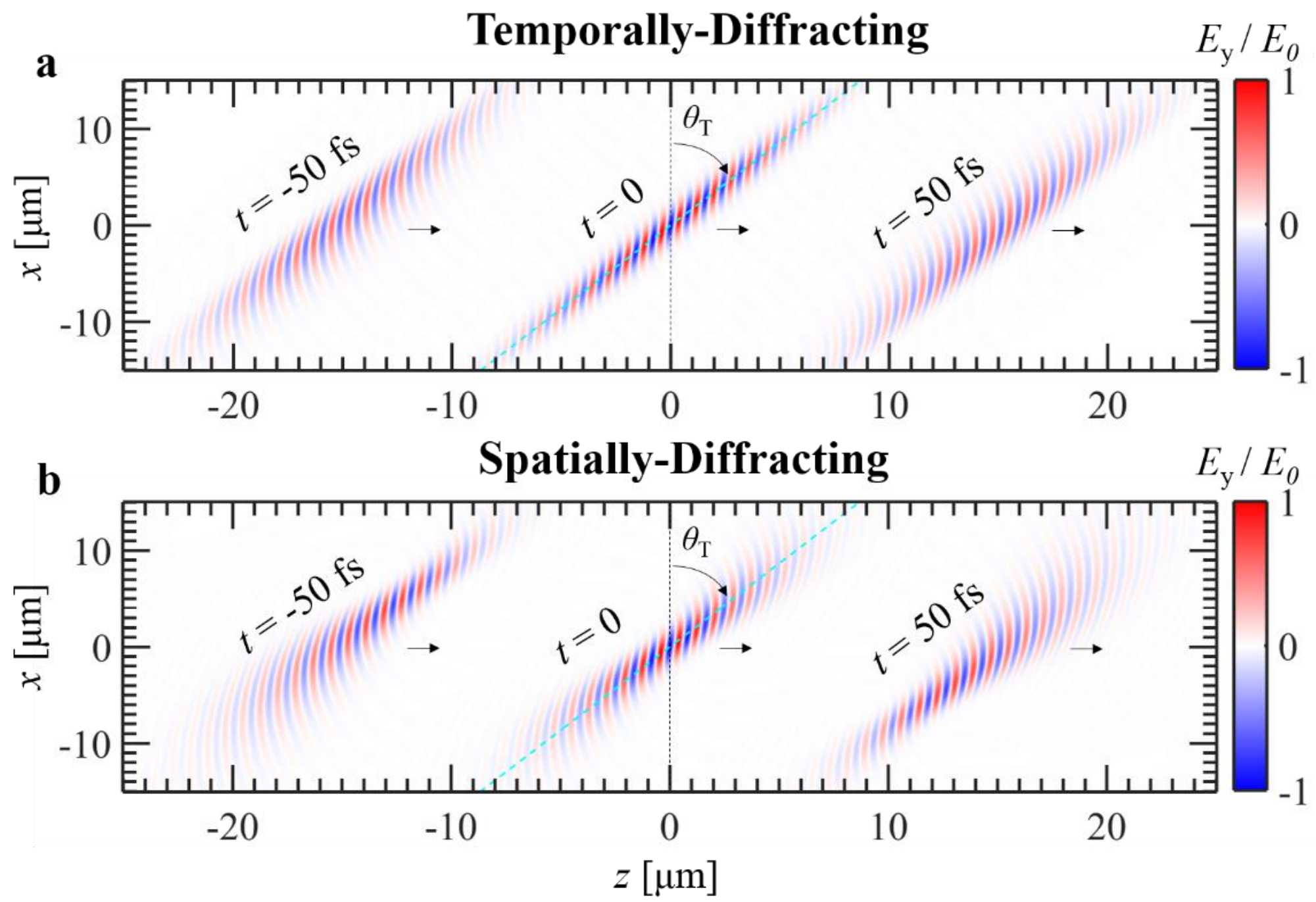

Figure 1. Propagation dynamics of ultrashort, nonparaxial tilted-pulse-front pulses. The temporally-diffracting pulse (a) and the spatially-diffracting pulse (b) shown correspond to exact, fully-closed form linearly-polarized electromagnetic pulses, obtained by substituting Eqs. (2) and (7) respectively into Eq. (1). The tilt angle $\theta_T$ is defined as the angle made by the pulse front (dotted cyan line) to the vertical axis. The phase and group velocities point in the z-direction (as indicated by black arrows). Here, $\theta_T = 30^o$, $k_0a = 19.7$ and $s = 911$. The peak wavelength $\lambda_0 = 0.8$ μm, i.e. peak frequency $\omega_0 = k_0c = 2\pi c/\lambda_0 = 2.36x10^{15}$ rad/s. $E_0$ is simply a normalizing constant. In this and other figures, the intensity profiles of the pulses follow the envelope of the plotted electric field profile without its carrier oscillations; we plot the field to fully depict the phase fronts.

The electromagnetic fields of a linearly-polarized pulse with a pulse-front-tilt angle of $\theta_T$, as shown in Fig. 1, are

$$\begin{aligned}\mathbf{E} &= \mathrm{Re}\left\{-\mu\frac{\partial}{\partial t}\nabla\times(\hat{x}\psi)\right\},\\ \mathbf{H} &= \mathrm{Re}\{\nabla\times\nabla\times(\hat{x}\psi)\}\end{aligned} \tag{1}$$

where $\mu$ is the permeability of the medium, $\hat{x}$ is the unit vector in the $x$ direction (resulting in an electric field that is primarily y-polarized), and $\psi$ is given as

$$\psi = \frac{1}{R}\left[1-\frac{\mathrm{i}\kappa_0}{s}(\zeta+\mathrm{i}R-\mathrm{i}a)\right]^{-s-1}, \tag{2}$$

and $R=\sqrt{\eta^2+y^2+(\mathrm{ic}\tau+a)^2}$ , $\kappa_0 = k_0|\sin\theta_T|$ , $\omega_0 = k_0\mathrm{c} = 2\pi\mathrm{c}/\lambda_0$ is the central angular frequency of the pulse, and c is the speed of light in the linear, homogeneous, isotropic, time-invariant medium. The parameters $a$ and $s$ control the width and pulse length [35]: $s$ is in general a complex number whose real part is positive, and $a$ is a positive real number. The variables $\tau$, $\eta$, $\zeta$ are defined as

$$\begin{aligned}\tau &= \frac{1}{\mathrm{c}|\sin\theta_T|}\left[\mathrm{c}t-(z-x\tan\theta_T)\cos^2\theta_T\right]\\ \eta &= \cot\theta_T\left(\mathrm{c}t-(z-x\tan\theta_T)\right)\\ \zeta &= |\sin\theta_T|(z+x\cot\theta_T)\end{aligned}. \tag{3}$$

It can be verified that $\psi$ exactly solves the scalar electromagnetic wave equation $\left(\nabla^2-1/\mathrm{c}^2\,\partial_t{}^2\right)\psi=0$, and consequently that the fields in Eq. (1) exactly solve the full Maxwell's equations.

To analyze the frequency-wavevector content of the electromagnetic pulse, we perform a four-dimensional Fourier transform of Eq. (2) to obtain

$$\Psi(\omega,k_x,k_y,k_z) = \Psi_0(\omega,k_x,k_z)\delta\left(\frac{\omega^2}{c^2} - k_x{}^2 - k_y{}^2 - k_z{}^2\right), \tag{4}$$

where the Dirac delta distribution $\delta$ restricts all components to lie on the surface of the light cone, as expected since the mode contains no evanescent wave components, and

$$\Psi_0(\omega,k_x,k_z) = \frac{4\pi}{c}\left(\frac{s}{\kappa_0}\right)^{s+1} \frac{K_z{}^s \exp(-sK_z/\kappa_0)}{\Gamma(s+1)} \exp\left[-a\left(\frac{\omega}{c} - k_z\right)\right]\theta\left(\frac{\omega}{c} - k_z\right)\theta(K_z), \tag{5}$$

where $\Gamma$ and $\theta$ are the Gamma and Heaviside functions respectively and $K_z = |\sin\theta_T|(k_z + k_x \cot\theta_T)$. Note from Eq. (5) that $\Psi_0 = 0$ when $K_z = 0$, and hence also when $\omega = 0$. This ensures the absence of zero-frequency components in Eqs. (1-3), making them an accurate description of a propagating pulse. Fig. 2a-c show the energy density spectra corresponding to the tilted-pulse-front pulse depicted in Fig. 1a. The energy density is defined as $U = 0.25\left[\varepsilon\left|\tilde{\mathbf{E}}(\omega,\mathbf{k})\right|^2 + \mu\left|\tilde{\mathbf{H}}(\omega,\mathbf{k})\right|^2\right]$, where $\tilde{\mathbf{E}}$ and $\tilde{\mathbf{H}}$ are respectively the Fourier-domain electric and magnetic fields readily obtained by substituting Eq. (5) into Eq. (1) in the Fourier domain.

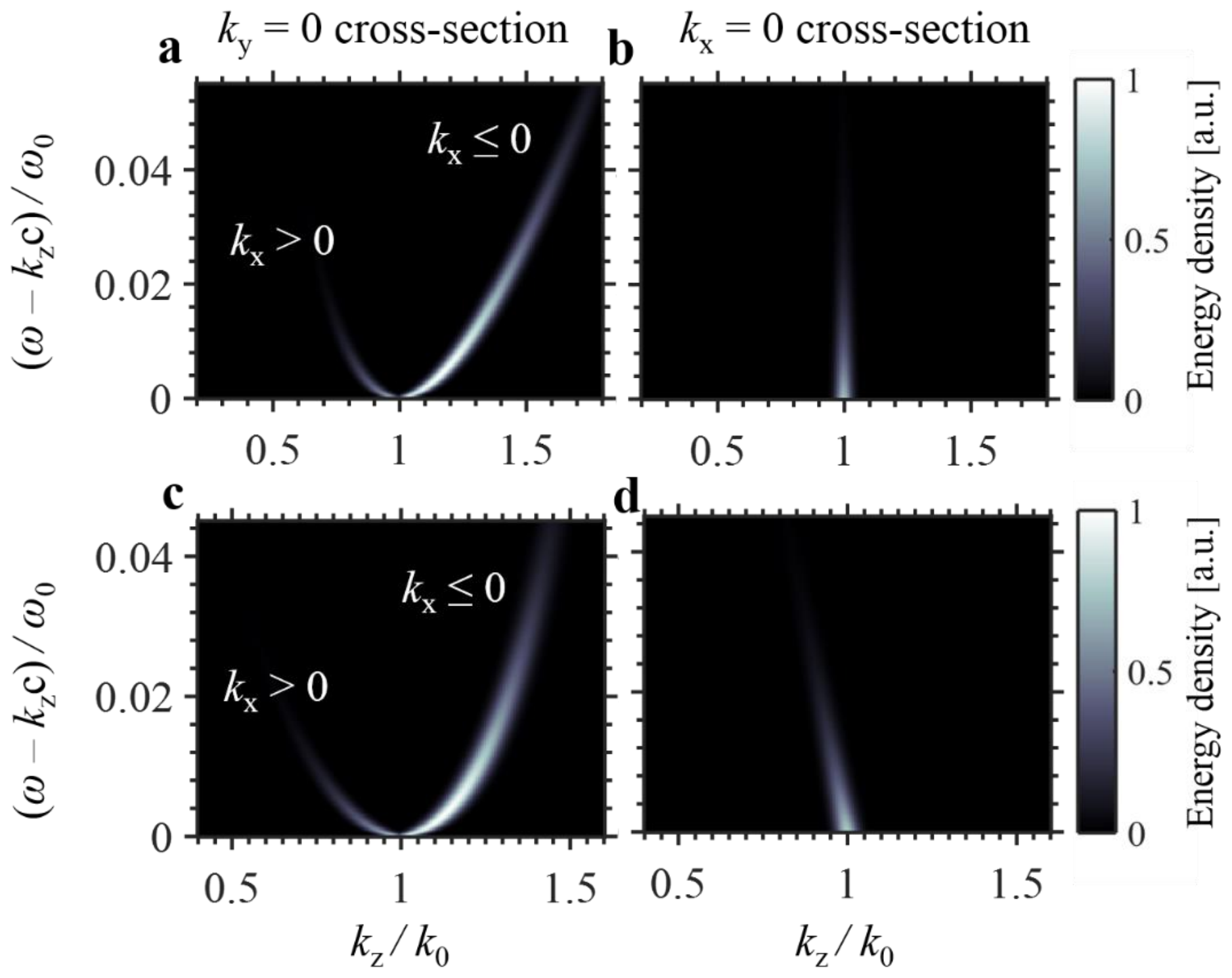

Figure 2. Spectral domain energy density distributions of temporally-diffracting (a-b) and spatially-diffracting (c-d) tilted-pulse-front pulses, corresponding to the pulses depicted in Fig. 1a and 1b respectively. These energy density distributions represent different cross-sections upon the surface of the light cone in four-dimensional frequency-wavevector space. In creating our wavepackets, the distributions in (a) and (c) have to include both the positive and the negative branches of $k_x = \pm\sqrt{\omega^2/c^2 - k_y{}^2 - k_z{}^2}$ (marked as $k_x \le 0$ and $k_x > 0$), satisfying the delta functions in Eqs. (4) and (8).

Although we present the y-polarized electromagnetic pulse as an example in Eq. (1) and Fig. 1, note that a wide variety of vectorial electromagnetic pulses with tilted pulse fronts may be readily generated from Eq. (2) using the Hertz vector potentials $\mathbf{\Pi}_e$ and $\mathbf{\Pi}_m$, via the equations [36]

$$\begin{aligned}\mathbf{E} &= \mathrm{Re}\left\{\nabla\times\nabla\times\mathbf{\Pi}_e - \mu\frac{\partial}{\partial t}\nabla\times\mathbf{\Pi}_m\right\} \\ \mathbf{H} &= \mathrm{Re}\left\{\nabla\times\nabla\times\mathbf{\Pi}_m + \varepsilon\frac{\partial}{\partial t}\nabla\times\mathbf{\Pi}_e\right\}\end{aligned}, \quad (6)$$

where $\varepsilon$ and $\mu$ are the medium's permittivity and permeability respectively. For instance, radially-polarized TM01 and azimuthally-polarized TE01 modes may be obtained by setting $\mathbf{\Pi}_\mathrm{e} = \psi\hat{\mathrm{z}}$, $\mathbf{\Pi}_\mathrm{m} = 0$, and $\mathbf{\Pi}_\mathrm{e} = 0$, $\mathbf{\Pi}_\mathrm{m} = \psi\hat{\mathrm{z}}$ respectively, whereas the linearly-polarized mode in x is obtained by setting $\mathbf{\Pi}_\mathrm{e} = \psi\hat{\mathrm{x}}$, $\mathbf{\Pi}_\mathrm{m} = 0$. Additionally, we note that Eq. (2) can be used to generate new classes of solutions (e.g., nonparaxial counterparts of the Hermite-Gaussian, Laguerre-Gaussian etc. families) since, for instance, any linear combination of any multiplicity of partial derivatives in space and time of (2) is also a solution of the scalar electromagnetic wave equation. Additional families are possible by substituting complex values for *s*.

Because Eq. (2) reduces to a temporally diffracting electromagnetic wavepacket [37] for the special case $\theta_\mathrm{T} = \pi/2$, we will refer to pulses obtained via Eq. (2) as temporally diffracting tilted-pulse-front pulses. We also obtain a second family of tilted-pulse-front pulses, distinct from those described by Eq. (2), as

$$\varphi = \frac{1}{2\mathrm{i}D}\left\{\left[1-\frac{\mathrm{i}\kappa_0}{s}\left(D-c\tau+\mathrm{i}a\right)\right]^{-s-1} - \left[1-\frac{\mathrm{i}\kappa_0}{s}\left(-D-c\tau+\mathrm{i}a\right)\right]^{-s-1}\right\}, \tag{7}$$

where $D = \sqrt{\eta^2 + y^2 + (\zeta - \mathrm{i}a)^2}$ and $\tau$, $\eta$, $\zeta$ are defined as in Eq. (3). Since Eq. (7) reduces to the laser pulse solution associated with the spatially complex source wave [38-40] for the special case $\theta_\mathrm{T} = \pi/2$, we refer to pulses obtained from Eq. (7) as spatially diffracting tilted-pulse-front pulses. By substituting $\psi$ for $\varphi$ in Eq. (1), we similarly obtain linearly-polarized spatially diffracting tilted-pulse-front pulses, which we compare with their temporally diffracting counterparts in Fig. 1.

Performing a four-dimensional Fourier transform of Eq. (7), we obtain

$$\Phi(\omega, k_x, k_y, k_z) = \Phi_0(\omega, k_x, k_z)\delta\left(\frac{\omega^2}{c^2} - k_x{}^2 - k_y{}^2 - k_z{}^2\right), \tag{8}$$

where

$$\begin{gathered}\Phi_0(\omega, k_x, k_z) = \frac{2\pi}{c}\left(\frac{s}{\kappa_0}\right)^{s+1} \frac{(\Omega/c)^s \exp(-s\Omega/\kappa_0 c)}{\Gamma(s+1)} \exp\left[-a\left(\frac{\omega}{c} - k_z\right)\right] \cdot \\ \left\{\theta\left(\frac{\omega}{c} - k_z\right) - \theta\left(-\frac{\omega}{c} - k_z\right)\right\}\theta(\Omega)\end{gathered}, \tag{9}$$

where $\Omega = \left[\omega - k_z c\cos^2\theta_T + k_x c\sin\theta_T\cos\theta_T\right] / \left|\sin\theta_T\right|$. As in the temporally diffracting case, we note the absence of zero-frequency and evanescent wave components, which make these expressions an accurate description of a propagating pulse. The spectra of the tilted-pulse-front pulse in Fig. 1b is shown in Fig. 2d-f.

From Fig. 1, we see that one prominent difference between the spatially and temporally diffracting versions for identical pulse parameters $a$, $s$, $\omega_0$ and $\theta_T$ is the variation in spatial curvature across different spatial cycles at a given instant in time. To further understand the nature of this difference, we examine $\psi$ and $\varphi$ in the paraxial, many-cycle limit i.e. $k_0 a >> 1$, $s >> 1$:

$$\psi \approx \psi_{\text{parax}} = \frac{1}{ic\tau + a}\exp\left[-\frac{\kappa_0(\eta^2 + y^2)}{2(ic\tau + a)}\right]\exp[i\kappa_0(z - ct)]\exp\left[-\frac{\kappa_0{}^2}{2s}(z - ct)^2\right], \tag{10}$$

$$\varphi \approx \varphi_{\text{parax}} = \frac{1}{2(i\zeta + a)}\exp\left[-\frac{\kappa_0(\eta^2 + y^2)}{2(i\zeta + a)}\right]\exp[i\kappa_0(z - ct)]\exp\left[-\frac{\kappa_0{}^2}{2s}(z - ct)^2\right]. \tag{11}$$

Equations (10) and (11) closely resemble the well-known Gaussian beam [41], being each composed of four factors: The first two being associated with the beam envelope, the third with the carrier and the last with the pulse envelope. The respective phases are given by

$$\angle\psi_{\text{parax}} = \kappa_0(z - \text{c}t) - \arctan(\text{c}\tau/a) + \frac{\kappa_0 \text{c}\tau(\eta^2 + y^2)}{2(\text{c}^2\tau^2 + a^2)}. \tag{12}$$

$$\angle\varphi_{\text{parax}} = \kappa_0(z - \text{c}t) - \arctan(\zeta/a) + \frac{\kappa_0 \zeta(\eta^2 + y^2)}{2(\zeta^2 + a^2)}. \tag{13}$$

Performing a coordinate rotation $x = x'\sin\theta_{\text{T}} + z'\cos\theta_{\text{T}}$, $z = -x'\cos\theta_{\text{T}} + z'\sin\theta_{\text{T}}$, we find that $\text{c}\tau = (\text{c}t + x'\cos\theta_{\text{T}})/|\sin\theta_{\text{T}}|, \eta = \cot\theta_{\text{T}}(\text{c}t + x'/\cos\theta_{\text{T}})$ and $\zeta = z'\text{sgn}(\sin\theta_{\text{T}})$: Notably, c$\tau$ and $\eta$ are independent of *z*', the coordinate directed along the pulse-front-tilt axis (dashed cyan line in Fig. 1). By making these substitutions, one can see from Eqs. (12) and (13) that only the third term in each case contributes to phase front curvature. However, this phase front curvature term varies as a function of z' only in the spatially diffracting case of Eq. (13). This agrees with what we see in Fig. 1, where the temporally diffracting pulse has a constant curvature across all spatial cycles at any instant in time, whereas the spatially diffracting pulse does not.

Associated with the temporally diffracting and spatially diffracting tilted-pulse-front solutions are their zero-bandwidth counterparts, which are obtained in the limit $s \to \infty$. Note that the term "zero-bandwidth" is used throughout this work to refer to the fact that the spectrum in the $\omega$-$k_z$ plane is given by a line of zero width; this can contain a range of frequencies and is different from the more common term "monochromatic" used for a single-frequency beam. These wavepackets are fascinating in their own right as nonparaxial nondiffracting modes of Maxwell's equations whose phase fronts are tilted with respect to the direction in which their intensity profile moves. Below, we describe the families of these nondiffracting wavepackets.

Fig. 3 presents wavepackets whose intensity profile moves with superluminal velocity $v_{\text{I}} = \beta_{\text{I}}$c ($|v_{\text{I}}| >$ c, $|\beta_{\text{I}}| > 1$) along the *z* direction and whose phase fronts are tilted by angle $\theta_{\text{P}}$ relative to the transverse (i.e. x-y) plane. Such wavepackets (which subsume the zero-bandwidth limit of the temporally diffracting pulses in Eq. (2)) are given by

$$\tilde{\psi} = \frac{1}{\tilde{R}} \exp\left[\tilde{\kappa}_0 \left(-\tilde{R} + a + i\tilde{\zeta}\right)\right], \tag{14}$$

where $\tilde{R} = \sqrt{\tilde{\eta}^2 + y^2 + \left(\mathrm{ic}\,\tilde{\tau} + a\right)^2}$ , $\tilde{\kappa}_0 = k_0 / \left|T_{11} - T_{31}\right|$ , $\mathrm{c}\tilde{\tau} = T_{11}\mathrm{c}t + T_{12}x + T_{13}z$ , $\tilde{\eta} = T_{21}\mathrm{c}t + T_{22}x + T_{23}z$ , $\tilde{\zeta} = T_{31}\mathrm{c}t + T_{32}x + T_{33}z$ , the subscripted *T* variables are elements of the matrix

$$\mathbf{T} = \begin{bmatrix} \sqrt{\dfrac{\beta_\mathrm{I}^2 + \tan^2\alpha}{\beta_\mathrm{I}^2 - 1}} & \dfrac{\tan\alpha}{\beta_\mathrm{I}} & -\dfrac{1}{\beta_\mathrm{I}}\sqrt{\dfrac{\beta_\mathrm{I}^2 + \tan^2\alpha}{\beta_\mathrm{I}^2 - 1}} \\ \chi\dfrac{\beta_\mathrm{I}\tan\alpha}{\left|\beta_\mathrm{I}\right|\sqrt{\beta_\mathrm{I}^2 - 1}} & \chi\dfrac{\sqrt{\beta_\mathrm{I}^2 + \tan^2\alpha}}{\left|\beta_\mathrm{I}\right|} & -\chi\dfrac{\tan\alpha}{\left|\beta_\mathrm{I}\right|\sqrt{\beta_\mathrm{I}^2 - 1}} \\ -\chi\dfrac{\beta_\mathrm{I}}{\left|\beta_\mathrm{I}\right|\sqrt{\beta_\mathrm{I}^2 - 1}} & 0 & \chi\dfrac{\left|\beta_\mathrm{I}\right|}{\sqrt{\beta_\mathrm{I}^2 - 1}} \end{bmatrix}, \tag{15}$$

$\chi = \mathrm{sgn}\left(\cos\alpha\right) = \mathrm{sgn}\left[\left(1 - \beta_\mathrm{I}\sec\theta_\mathrm{P}\right)/\left(\sec\theta_\mathrm{P} - \beta_\mathrm{I}\right)\right]$, $\alpha = \alpha_B + \pi\theta\left(-\chi\right)$ and

$$\alpha_\mathrm{B} = \arctan\left[\frac{-\tan\theta_\mathrm{P}\left|\beta_\mathrm{I}\right|\sqrt{\beta_\mathrm{I}^2 - 1}}{\sec\theta_\mathrm{P} - \beta_\mathrm{I}}\chi\right], \tag{16}$$

where $-\pi/2 \le \alpha_\mathrm{B} \le \pi/2$. Substituting Eq. (14) into Eq. (1) with $\psi$ replaced by $\tilde{\psi}$ gives vectorial electromagnetic fields that are exact solutions to Maxwell's equations.

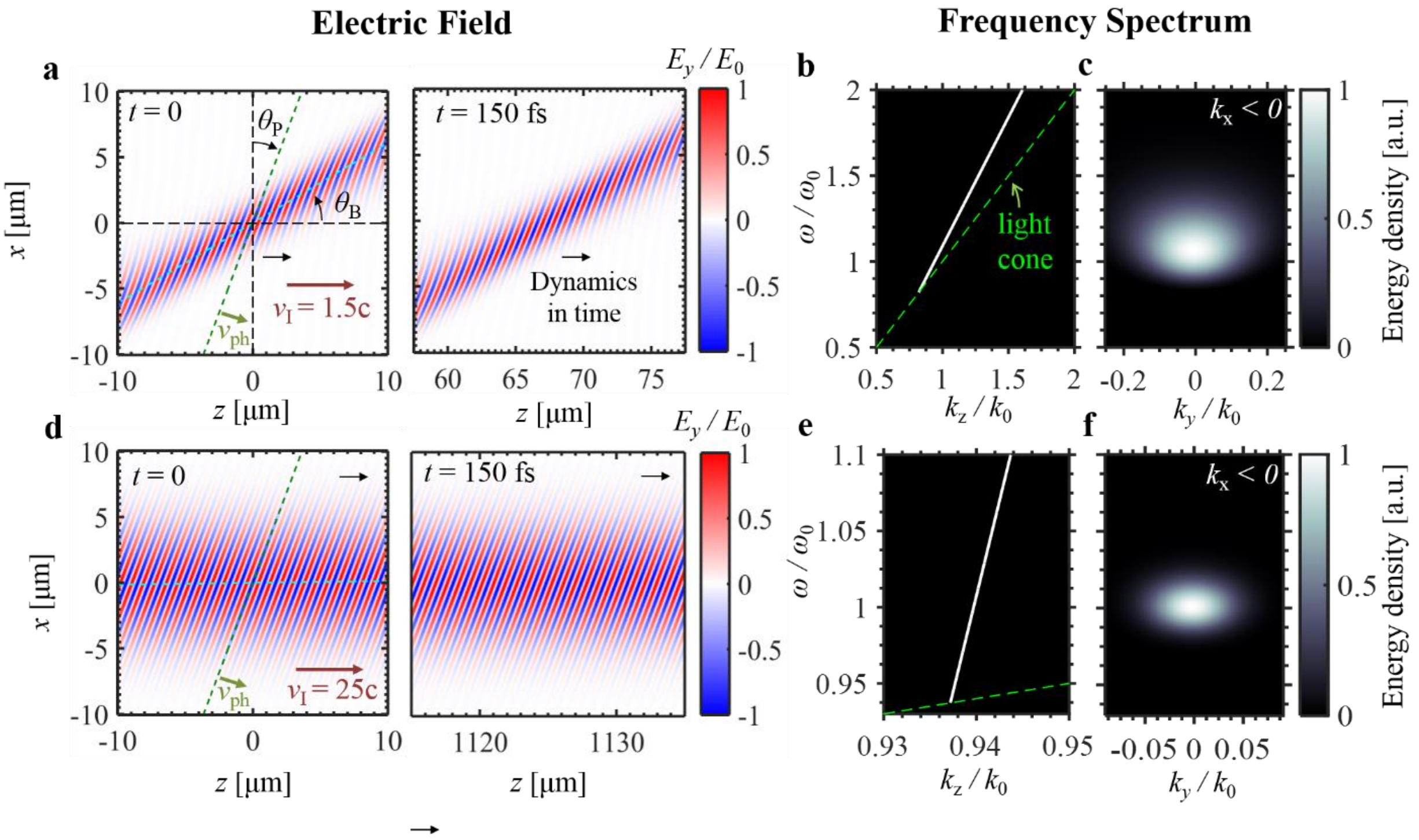

Figure 3. Nondiffracting wavepackets with intensity profiles that travel at superluminal speeds $v_I$ = 1.5c (a-c) and $v_I$ = 25c (d-f) in the z direction (c being the speed of light). The phase-front-tilt angle $\theta_P$ is defined by the angle the phase front (dashed green lines in a and d) makes with the vertical, whereas the beam-tilt angle $\theta_B$ is the angle the dashed cyan in a and d line makes with horizontal. The phase front propagates at phase velocity $v_{ph}$ that is in general different in direction and magnitude from envelope velocity $v_I$. The wavepackets in a and d contain frequency components only along the white line in b and e respectively. Note that the white line lies entirely within the light cone in each case since the wavepackets contain no evanescent components. As can be seen from the spectral energy density in c and f for their respective cases, the largest frequency components occur at the peak frequency $\omega_0 = k_0 c = 2\pi c/\lambda_0 = 2.36\times10^{15}$ rad/s. $E_0$ is simply a normalizing constant. In both cases, $\theta_P = 20^o$. $k_0 a$ = 19.7 in (a-c), whereas $k_0 a$ = 493 in (d-f). Importantly, note that while the intensity patterns in (a) and (d) at any instant in time resemble those of a diffracting beam, their nondiffracting natures lies in the fact that these patterns translate perfectly in $z$ as a function of time.

The angle $\theta_B$ in Fig. 3 is automatically fixed by the choice of $v_I$ and $\theta_P$ as $\theta_B = \arctan\left\{\tan\alpha \Big/ \left(\sqrt{\beta_I^2 - 1}\sqrt{\beta_I^2 + \tan^2\alpha}\right)\right\}$. It is interesting to note that the tilt of the beam $\theta_B$ and the tilt of the phase front $\theta_P$ are in general not orthogonal to each other. Additionally, the phase velocity (directed perpendicular to the phase front) does not point in the same direction as the intensity peak velocity in general. The group velocity is only defined in case of finite bandwidth (as in Figs.1 and 2), and is always parallel to the phase velocity. Finite-energy wavepackets may be obtained by integrating Eq. (14) over $\tilde{\kappa}_0$, constructing the finite-bandwidth wavepacket from a superposition of zero-bandwidth wavepackets. The result is a diffracting pulse whose group velocity points in the same direction as the phase velocity, and whose intensity peak shifts in time (at an approximate speed of $v_{\mathrm{I}}$, for a small bandwidth of integration) within the pulse.

In the frequency-wavevector domain, we find that Eq. (14) corresponds to the expression

$$\tilde{\Psi}(\omega, k_{\mathrm{x}}, k_{\mathrm{y}}, k_{\mathrm{z}}) = \tilde{\Psi}_0(\omega, k_{\mathrm{x}}, k_{\mathrm{z}})\delta\left(\frac{\omega^2}{\mathrm{c}^2} - k_{\mathrm{x}}{}^2 - k_{\mathrm{y}}{}^2 - k_{\mathrm{z}}{}^2\right), \tag{17}$$

where

$$\tilde{\Psi}_0(\omega, k_{\mathrm{x}}, k_{\mathrm{z}}) = \frac{4\pi}{\mathrm{c}}\delta\left(\tilde{K}_z - \tilde{\kappa}_0\right)\exp\left[-a\left(\frac{\tilde{\Omega}}{\mathrm{c}} - \tilde{\kappa}_0\right)\right]\theta\left(\frac{\tilde{\Omega}}{\mathrm{c}} - \tilde{\kappa}_0\right), \tag{18}$$

$\tilde{K}_z = T_{31}\,\omega/\mathrm{c} + T_{32}k_{\mathrm{x}} + T_{33}k_{\mathrm{z}}$ and $\tilde{\Omega} = T_{11}\omega + T_{12}k_{\mathrm{x}}\mathrm{c} + T_{13}k_{\mathrm{z}}\mathrm{c}$ .

Fig.4 presents wavepackets whose intensity profile moves with subluminal velocity $v_{\mathrm{I}} = \beta_{\mathrm{I}}\mathrm{c}$ ($|v_{\mathrm{I}}| < \mathrm{c}$, $|\beta_{\mathrm{I}}| < 1$) along the $z$ direction and whose phase fronts are tilted by angle $\theta_{\mathrm{P}}$ relative to the transverse (i.e. x-y) plane. Such wavepackets (which subsume the zero-bandwidth limit of the spatially diffracting pulses in Eq. (7)) are given by

$$\tilde{\varphi} = \frac{1}{\tilde{D}}\sin\left(\tilde{\kappa}_0{}'\tilde{D}\right)\exp\left[-\tilde{\kappa}_0{}'\left(a + \mathrm{c}\tilde{\tau}'\right)\right], \tag{19}$$

where $\tilde{D} = \sqrt{\tilde{\eta}'^2 + y^2 + \left(\mathrm{ic}\,\tilde{\tau}' + a\right)^2}$ , $\tilde{\kappa}_0{}' = k_0 / \left|U_{11} - U_{31}\right|$ , $\mathrm{c}\tilde{\tau}' = U_{11}\mathrm{c}t + U_{12}x + U_{13}z$ , $\tilde{\eta}' = U_{21}\mathrm{c}t + U_{22}x + U_{23}z$, $\tilde{\zeta}' = U_{31}\mathrm{c}t + U_{32}x + U_{33}z$, the subscripted $U$ variables are elements of the matrix

$$\mathbf{U}=\begin{bmatrix} \frac{1}{\sqrt{1-\beta_{\mathrm{I}}^{2}}} & 0 & -\frac{\beta_{\mathrm{I}}}{\sqrt{1-\beta_{\mathrm{I}}^{2}}} \\ \frac{\beta_{\mathrm{I}}}{\sqrt{1-\beta_{\mathrm{I}}^{2}}}\sin\alpha' & \cos\alpha' & -\frac{1}{\sqrt{1-\beta_{\mathrm{I}}^{2}}}\sin\alpha' \\ -\frac{\beta_{\mathrm{I}}}{\sqrt{1-\beta_{\mathrm{I}}^{2}}}\cos\alpha' & \sin\alpha' & \frac{1}{\sqrt{1-\beta_{\mathrm{I}}^{2}}}\cos\alpha' \end{bmatrix}, \qquad (20)$$

$\alpha'=\alpha_B'+\pi\theta(-\chi')$, $\chi'=\mathrm{sgn}(\cos\alpha')=\mathrm{sgn}\left((\beta_{\mathrm{I}}-\sec\theta_{\mathrm{P}})/(\beta_{\mathrm{I}}\sec\theta_{\mathrm{P}}-1)\right)$ and

$$\alpha_{\mathrm{B}}'=\arctan\left[\frac{\tan\theta_{\mathrm{P}}\sqrt{1-\beta_{\mathrm{I}}^{2}}}{\beta_{\mathrm{I}}\sec\theta_{\mathrm{P}}-1}\right], \qquad (21),$$

where $-\pi/2\le\alpha_{\mathrm{B}}'\le\pi/2$. Substituting Eq. (19) into Eq. (1) with $\psi$ replaced by $\tilde{\varphi}$ gives vectorial electromagnetic fields that are exact solutions to Maxwell's equations.

The angle $\theta_{\mathrm{B}}$ in Fig. 4 is automatically fixed by the choice of $v_{\mathrm{I}}$ and $\theta_{\mathrm{P}}$ as $\theta_{\mathrm{B}}=\arctan\left\{\tan\alpha/\sqrt{1-\beta_{\mathrm{I}}^{2}}\right\}$. As in the temporally diffracting case, $\theta_{\mathrm{B}}$ and the tilt of the phase front $\theta_{\mathrm{P}}$ are in general not orthogonal to each other, and finite-energy wavepackets may be obtained by integrating Eq. (19) over $\tilde{\kappa}_0'$. In the frequency-wavevector domain, we find that Eq. (19) corresponds to the expression

$$\tilde{\Phi}(\omega,k_{\mathrm{x}},k_{\mathrm{y}},k_{\mathrm{z}})=\tilde{\Phi}_0(\omega,k_{\mathrm{x}},k_{\mathrm{z}})\delta\left(\frac{\omega^2}{\mathrm{c}^2}-k_{\mathrm{x}}^{2}-k_{\mathrm{y}}^{2}-k_{\mathrm{z}}^{2}\right), \qquad (22)$$

where

$$\tilde{\Phi}_0(\omega,k_{\mathrm{x}},k_{\mathrm{z}})=2\pi\delta\left(\tilde{\Omega}'-\tilde{\kappa}_0'\mathrm{c}\right)\exp\left[-a\left(\tilde{\kappa}_0'-\tilde{K}_z'\right)\right]\left[\theta\left(\tilde{\kappa}_0'-\tilde{K}_z'\right)-\theta\left(-\tilde{\kappa}_0'-\tilde{K}_z'\right)\right], \qquad (23)$$

$\tilde{K}_z'=U_{31}\,\omega/\mathrm{c}+U_{32}k_{\mathrm{x}}+U_{33}k_{\mathrm{z}}$ and $\tilde{\Omega}'=U_{11}\omega+U_{12}k_{\mathrm{x}}\mathrm{c}+U_{13}k_{\mathrm{z}}\mathrm{c}$.

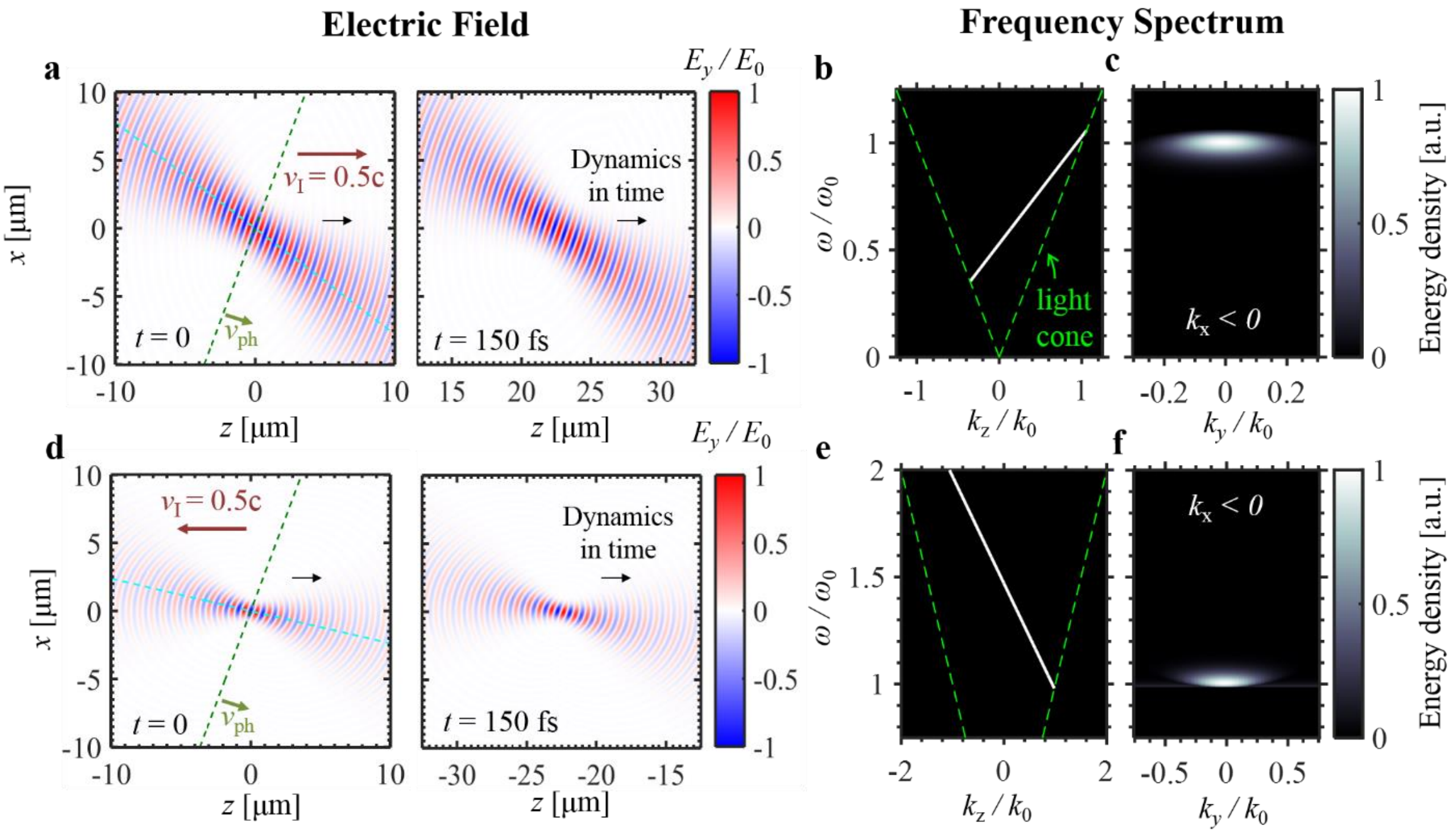

Figure 4. Nonparaxial nondiffracting wavepackets with intensity profiles that travel at subluminal speed $v_I = 0.5c$ in the forward (a-c) backward (d-f) direction along the z axis (c being the speed of light). Dashed green and cyan lines are as described in the caption of Fig. 3. The phase front propagates at phase velocity $v_{ph}$ that is in general different in direction and magnitude from envelope velocity $v_I$. The wavepackets in a and d contain frequency components only along the white line in b and e respectively. Note that the white line lies entirely within the light cone in each case since the wavepackets contain no evanescent components. As can be seen from the spectral energy density in c and f for their respective cases, the largest frequency components occur at the peak frequency $\omega_0 = k_0c = 2\pi c/\lambda_0 = 2.36\text{x}10^{15}$ rad/s. In both cases, $\theta_P = 20^o$ and $k_0a = 19.7$. Importantly, note that while the intensity patterns in (a) and (d) at any instant in time resemble those of a diffracting beam, their nondiffracting natures lies in the fact that these patterns translate perfectly in $z$ as a function of time.

It is interesting to note that setting $\theta_P = 0$ in Eq.(19) reduces it to the subluminal MacKinnon-type pulse (Eq. (8) of ref. [24]). In contrast, setting $\theta_P = 0$ in Eq.(14) reduces it to the superluminal focus wave mode (Eqs. (10) and (11) of ref. [24], for intensity peak velocity $v_I < 0$ and $v_I > 0$ respectively). Additionally, we note that our tilted-pulse-front pulses and nondiffracting tilted-phase-front wavepackets – which subsume the focus wave modes [20-24] – can be obtained from the temporally complex source wave [37] and the spatially complex source wave [38-40] via a Lorentz boost. More generally, given an expression $\psi(x^\mu)$ – $x^\mu$ being the position four-vector (c$t$,$x$,$y$,$z$) indexed by $\mu$ – that exactly solves the electromagnetic wave equation, a family of exact wavepacket solutions is given by $\psi(\Lambda^\mu{}_\nu x^\nu)$, where $\Lambda^\mu{}_\nu$ is an arbitrary Lorentz boost tensor. The four-dimensional Fourier transform of $\psi(\Lambda^\mu{}_\nu x^\nu)$ is simply $\Psi(\Lambda^\mu{}_\nu k^\nu)$, where $\Psi(k^\mu)$ is the Fourier transform of $\psi(x^\mu)$.

The electromagnetic wavepackets presented above are exact solutions in a dispersion-free medium. However, for applications like terahertz generation, second harmonic generation, and optical parametric amplification, one may be interested in the propagation of such electromagnetic wavepackets inside a dispersive medium. Their evolution in the presence of dispersion may also be analytically modelled based on these equations by substituting the speed of light c with an expression that changes with frequency

$$\mathrm{c} \rightarrow \mathrm{c} \Bigg/ \left[ 1 + \frac{n'}{n_0}(\omega - \omega_0) + \frac{n''}{n_0}(\omega - \omega_0)^2 \right], \tag{24}$$

where c on the right-hand-side now denotes the speed of light in the medium at frequency $\omega_0$, whereas $n_0$, n’ and n” are respectively the zeroth, first and second derivative with respect to angular frequency of the medium index, evaluated at frequency $\omega_0$. By performing a Taylor expansion of its frequency domain expression and retaining terms up to the second power of $(\omega - \omega_0)$, we obtain the dispersive version of temporally-diffracting tilted-pulse-front wavepacket $\psi$ in Eq. (2) in the space-time domain as

$$\begin{aligned} \psi_{\mathrm{disp}} \approx\ & \psi - \mathrm{i}\frac{n'}{n_0}(1 - k_0 a)\exp(-\mathrm{i}\omega_{0\text{-}} t)\frac{\partial}{\partial t}[\psi \exp(\mathrm{i}\omega_{0\text{-}} t)] \\ & - \frac{1}{2}\left[ k_0 a (k_0 a - 2)\frac{n'^2}{n_0{}^2} - \frac{2a}{\mathrm{c}}\frac{n'}{n_0} + (1 - k_0 a)\frac{n''}{n_0} \right]\exp(-\mathrm{i}\omega_{0\text{-}} t)\frac{\partial^2}{\partial t^2}[\psi \exp(\mathrm{i}\omega_{0\text{-}} t)] \end{aligned} . \tag{25}$$

Through the same procedure, we obtain the dispersive version of the spatially-diffracting tilted-pulse-front wavepacket $\varphi$ in Eq. (7) as

$$\begin{aligned}\varphi_{\mathrm{disp}} \approx \varphi - \mathrm{i}\frac{n'}{n_0}\left[s+1-k_0 b\right]\exp\left(-\mathrm{i}\omega_{0\text{-}}t\right)\frac{\partial}{\partial t}\left[\varphi\exp\left(\mathrm{i}\omega_{0\text{-}}t\right)\right] \\ -\frac{1}{2}\left[k_0 b\left(k_0 b+(s-2)(s+1)\right)\frac{n'^2}{n_0^{\;2}} - \frac{2}{\mathrm{c}}b\frac{n'}{n_0} + \left(s+1-k_0 b\right)\frac{n''}{n_0}\right]\cdot, \\ \exp\left(-\mathrm{i}\omega_{0\text{-}}t\right)\frac{\partial^2}{\partial t^2}\left[\varphi\exp\left(\mathrm{i}\omega_{0\text{-}}t\right)\right]\end{aligned} \tag{26}$$

where $b \equiv s/\left(\kappa_0\left|\sin\theta_{\mathrm{T}}\right|\right)+a$. Similarly, we obtain the dispersive version of the superluminal tilted-phase-front nondiffracting beam $\tilde{\psi}$ in Eq. (14) as

$$\begin{aligned}\tilde{\psi}_{\mathrm{disp}} \approx \tilde{\psi} - \mathrm{i}\frac{n'}{n_0}\left(1-k_0 d\right)\exp\left(-\mathrm{i}\omega_{0\text{-}}t\right)\frac{\partial}{\partial t}\left[\tilde{\psi}\exp\left(\mathrm{i}\omega_{0\text{-}}t\right)\right] \\ -\frac{1}{2}\left[k_0 d\left(k_0 d-2\right)\frac{n'^2}{n_0^{\;2}} - \frac{2d}{\mathrm{c}}\frac{n'}{n_0} + \left(1-k_0 d\right)\frac{n''}{n_0}\right]\exp\left(-\mathrm{i}\omega_{0\text{-}}t\right)\frac{\partial^2}{\partial t^2}\left[\tilde{\psi}\exp\left(\mathrm{i}\omega_{0\text{-}}t\right)\right],\end{aligned} \tag{27}$$

where $d \equiv a\left(T_{11}-T_{31}\right)$, and the dispersive version of the subluminal tilted-phase-front nondiffracting beam $\tilde{\varphi}$ in Eq. (19) as

$$\begin{aligned}\tilde{\varphi}_{\mathrm{disp}} \approx \tilde{\varphi} - \mathrm{i}\frac{n'}{n_0}k_0 f\exp\left(-\mathrm{i}\omega_{0\text{-}}t\right)\frac{\partial}{\partial t}\left[\tilde{\varphi}\exp\left(\mathrm{i}\omega_{0\text{-}}t\right)\right] \\ -\frac{1}{2}\left[k_0^{\;2} f^{\,2}\frac{n'^2}{n_0^{\;2}} - \frac{2f}{\mathrm{c}}\frac{n'}{n_0} - k_0 f\frac{n''}{n_0}\right]\exp\left(-\mathrm{i}\omega_{0\text{-}}t\right)\frac{\partial^2}{\partial t^2}\left[\tilde{\varphi}\exp\left(\mathrm{i}\omega_{0\text{-}}t\right)\right],\end{aligned} \tag{28}$$

where $f \equiv a\left(U_{11}-U_{31}\right)$. Note that the vectorial electromagnetic fields are readily obtained from the foregoing scalar expressions using the Hertz potentials Eq. (6), which reduces to Eq. (1) in the linearly-polarized case.

The nonparaxial tilted-pulse-front pulses we present here are potentially useful in applications like electron acceleration [6,7] and X-ray generation [9,10], where the use of tightly-focused, few-cycle driving laser pulses can achieve larger intensities and strong-field dynamics with a smaller amount of energy. The nondiffracting tilted-phase-front beams we introduce potentially open up new options for phase-matching in terahertz generation. For instance, the intensity peak velocity of a subluminal nondiffracting optical beam (Eqs. (19-21)) could be set equal to the speed of terahertz phase propagation in the generating medium (e.g., Lithium Niobate) under a collinear phase-matching scheme. Alternatively, it may be possible to use the superluminal nondiffracting beam (Eq. (14)) by choosing $v_I$ and $\theta_B$ such that the phase-matching condition is satisfied.

Tilted-pulse-front pulses can be generated through the use of diffraction gratings or prisms to introduce angular dispersion, or through the imposition of spatial chirp [42, 43]. These techniques, which have been applied towards generating pulses in the paraxial regime, can potentially be extended to the few-cycle, nonparaxial regime by identifying materials with a suitably broadband response. Besides this, the use of pre-engineered phase masks and amplitude masks or spatial light modulators (SLMs), in either real space or Fourier space, could be a viable option for experimentally realizing the wavepackets presented here at optical frequencies [44]. Importantly, one can shape nonparaxial wavepackets by performing the shaping in the paraxial domain (e.g., with regular phase masks), before strongly focusing the result with a high-NA objective. This requires design of the paraxial shaping stage to compensate for the focusing stage so the final focused beam has the intended profile, a process that must be guided by an accurate nonparaxial model for the electromagnetic wavepacket. This approach has been experimentally applied, for example, in the realization of nonparaxial accelerating beams [32, 45]. Shaping of the spectrum in the $k_z$ and $\omega$ dimensions (tantamount to temporal pulse shaping) may be achieved by using a prism or grating to convert the $k_z$ and $\omega$ parameters to spatial dimensions and shaping them directly with conventional light-shaping elements. The range of alternative approaches continues to be broadened by the discovery of new spatiotemporal pulse shaping methods [46].

Yet another possible realization would be through time-dependent current distributions located far from the respective focal regions, which would generate the wavepackets studied in Figs. 1-4. Specifically, one can compute a localized current distribution via the expression $\mathbf{J}_s = \hat{z} \times \mathbf{H}$, where Re{$\mathbf{J}_s$} is the current distribution in the x-y plane at a given $z$. The radiation from such a current distribution will then produce the electromagnetic wavepacket whose magnetic field is described by **H**. It may be possible to induce such current density modulations using an ordinary pulsed laser incident on a metasurface, where nano-antenna arrays are used to create the desired current distributions.

We have presented exact, analytical expressions for the electromagnetic fields of tilted-pulse-front pulses, which encompass conditions of extreme nonparaxiality and extremely short pulse duration. We have shown that their zero-bandwidth counterparts are fascinating in their own right as new classes of nonparaxial nondiffracting electromagnetic modes. The intensity envelopes of these wavepackets move at a constant velocity that can be much greater than or much less than the speed of light, and generally travel at an angle (including backwards) with respect to the direction of phase front propagation. The electromagnetic modes we present are closed-form solutions to Maxwell's equations, which makes them attractive candidates for modeling nonparaxial pulses in light-matter interaction scenarios ranging from particle acceleration and X-ray generation to optical parametric amplification and materials processing.

**Funding Information:** The work was supported by the Science and Engineering Research Council (SERC; grant no. 1426500054) of the Agency for Science, Technology and Research (A*STAR), Singapore. The research of I.K. was supported by the Seventh Framework Programme of the European Research Council (FP7–Marie Curie IOF) under grant agreement no. 328853 – MC–BSiCS.

**Acknowledgments:** N.A.